\documentclass{kluwer}
\usepackage{epsfig}
\newdisplay{guess}{Conjecture}

\begin{document}


\def\b12{PSR B1259-63}
\def\beq{\begin{equation}}
\def\eeq{\end{equation}}
\def\a{\alpha}
\def\g{\gamma}
\def\s{\sigma}
\def\e{\varepsilon}
\def\th{\theta}
\def\om{\omega}
\renewcommand{\deg}{$^{\circ}$}
\newcommand{\degm}{^{\circ}}
\newcommand{\etal}{{et~al.$\,$}}

\begin{opening}
\title{Non-pulsed emission\\ from the binary system \b12 }
\author{Maria \surname{Chernyakova}}
\author{Andrei \surname{Illarionov}}
\runningauthor{Maria A. Chernyakova, Andrei F. Illarionov}
\runningtitle{Non-pulsed emission from the binary system \b12}
\institute{Astro Space Center of P.N.Lebedev Physical Institute}

\begin{abstract}
\b12 is the only known binary system with a radio pulsar from which the
non-pulsed radio and X-ray emission was detected. The companion star in this
system is a Be star SS 2883. A rapidly rotating radio pulsar is expected to
produce a wind of relativistic
particles. Be stars are known to produce highly asymmetric mass loss. Due to
the interaction of the pulsar wind and the Be star wind the system of
two shocks  between the pulsar and the Be star forms. In this paper we show
that the observed non-pulsed radio emission from the system is a result of
the synchrotron emission of the relativistic particles in the outflow
beyond the shock wave and that the non-pulsed X-ray emission is due to the
inverse Compton scattering of the Be star photons on this
particles.  \end{abstract} \keywords{pulsars:individual:\b12}

\end{opening}

{\small
\section{Introduction}
\b12 is the only known binary system with a radio pulsar from which the
non-pulsed radio and X-ray emission was detected. The radio pulsar \b12
of spin period $P$=47.76ms was discovered in a high-frequency survey of
the southern Galactic plane.  Spindown luminosity is
equal to $L_p=9\times10^{35}$erg/s. Pulsar timing confirmed by later
observations revealed that the pulsar is in a high-eccentricity ($e$=0.87),
long-orbital period ($T$=3.4year) binary system with a massive companion.
  Its companion  is a 10th mag Be star SS2883 of luminosity 
$L_*$=5.8$\times$10$^4L_\odot$, estimated radius
$R_*\sim$(6-10)$R_\odot$, mass $M\sim$10$M_\odot$. The distance to the
Earth is about  2 kpc,  the binary separation at
periastron is equal to 10$^{13}$cm.
 Assuming a pulsar mass of 1.4$M_\odot$ the implied inclination angle of 
the binary orbit to the plane of sky is 36$^\circ$. The longitude of periastron is equal to 138$^\circ$ (Johnston \etal 1996).  Timing measurements have
 shown that the disc of the Be star 
is likely to be highly inclined to the orbital plane (Wex \etal, 1988).

The extensive radio observations of \b12  were carried out during the
1994 and 1997 periastron passage at frequencies 1.5, 2.3, 4.8 and 8.4 GHz
(Johnston \etal 1996, 1999).  It turned out that when the pulsar is far
from the periastron the pulsar emission is highly linear polarized and its
intensity is practically independent on the pulsar orbital position.  But
as the pulsar approaches to the periastron the properties of the pulsar
emission start to change.  The depolarization of pulsed emission, the
increase of the dispersion measure and the absolute value of the rotation
measure near the periastron occurs. The pulsar was not detected in 1.5 GHz
 data on 1993 December 20 (20 days before the periastron) and reappeared only on
1994 February 4 (24 days after the periastron).  During the 44 days eclipse no
pulse emission was detected in extensive observations at 1.5 and 8.4 GHz.
 In the works of Lipunov \etal (1994)
and Johnston \etal (1996) it was shown that the observed eclipse is due to
the free-free absorption of the pulsar emission in the Be star disk.

During the
1994 and 1997 periastron passage the observations of the unpulsed
radio emission coming from the system were made using the Molongo Observatory
Syn\-thesis Telescope (MOST) and the Australia Telescope Compact Array
(ATCA) at five frequencies between 0.84 and 8.4 GHz (Johnston \etal 1999).
The first significant data at all frequencies were received 21 day before
the periastron.  Since that time  the intensity of the non-pulsed emission by
and large gradually increases  until the 20th day after the periastron. That
day it reaches its maximum (57 mJy at 1.4 GHz) and then it starts to
decline.  The source remained significant in the ATCA data until about 60
days after the periastron and until about hundred days after the periastron
in the MOST data. At times when the pulsar emission is eclipsed, there is
no evidence of polarized emission in any of the observations in either 1994
or 1997.

\begin{figure} 
\centering\epsfig{file=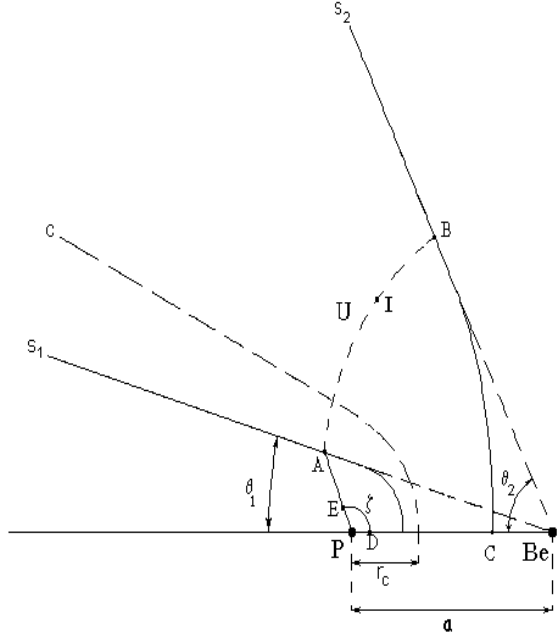,width=12cm, height=20cm}
\vspace{-10cm}
\caption {The geometry under consideration. Point $P$ marks the position
of the pulsar, point $Be$ - the position  of the Be star. Curves $S_1$,
$S_2$ represent the shock waves, curve $c$ - the contact surface. Curve
$U$ is a curve of the constant density (see text). Points $A$ and $B$ mark
the intersection of the curve $U$ with the shocks $S_1$ and $S_2$,
point $C$ is the point of the shock $S_2$ intersection with the binary
axis.  $ED$ is a small arc centred at the point $P$.}
\end{figure}

Since its discovery the PSR B1259-63 system was observed several
times by  X-ray instruments
(see review in Tavani \etal 1997, Hirayama \etal 1999).
It was shown that
the X-ray spectrum of the system is consistent with a power-law of photon
index $\sim-1.7$. No significant X-ray pulsations with the pulsar spin
period were detected.

 The aim of this paper is to explain the origin of the non-pulsed radio and
X-ray emission. It is well-known that pulsars lose rotational energy in
the form of relativistic MHD winds (Michel 1969, Arons 1992). The wind
carries energy flux in the form of electromagnetic fields and the kinetic
energy of the relativistic electrons and positrons.  The energy flux in the
pulsar wind ${\cal F}_w$ at distance $r$ well over the light cylinder is equal  to
\begin{equation}
 {\cal F}_w=\frac{L_p}{4\pi r^2}=\frac{cB_w^2}{4\pi\sigma}.
\label{b0}
 \end{equation}
 $B_w$ is a magnetic field in the wind, $\sigma\ll 1$
is a pulsar wind magnetization, i.e. the upstream ratio of electromagnetic
 energy density and particle kinetic energy density. For the Crab Nebula
 $\sigma\sim0.005$ (Kennel\& Coronity, 1984).

Be stars are well-known to be the source of the strong highly
anisotropic matter outflow. Both a dilute polar wind and a denser
equatorial disk have been invoked to reconcile models for IR, UV and
optical observations of Be stars (Waters \etal 1988).
 Due
to the interaction of the pulsar and Be-star winds the system of two shocks arises and
the relativistic particles flowing at first radially from the pulsar turn
after passing the shock and start to outflow along the contact surface.
In the paper of Tavani \etal (1997) the non-pulsed X-ray emission was explained as a result of the synchrotron emission of the highly relativistic pulsar wind particles with Lorentz factor $\gamma \sim 10^6$. In this model the origin of the unpulsed radio emission is unclear.  Here we propose a model in which Lorentz factor is supposed to be much more moderate, $\gamma \sim 10 $, and show that in this case the unpulsed radio emission is a result of the
synchrotron radiation of the relativistic particles in the outflow beyond
the shock and that the non-pulsed X-ray emission is due to the
inverse Compton scattering of the Be star photons on this particles.

\section{Colliding winds}

In order to calculate the unpulsed emission coming from \b12
it is necessary to know
the structure of the shock waves arising in the system due to the interaction
of the relativistic pulsar wind and a non-relativistic Be star wind.

The structure of the colliding winds in the system \b12 is very complicate.
But  the origin of the unpulsed emission can be understood with 
the help of the rather simple model.
  For the simplicity we suppose that the polar wind of the
Be-star occupies all the region outside the disk and that in that region it
is homogeneous. The pulsar wind we suppose to be spherically symmetric. The
observations of the inner regions of the Crab Nebula (Hester \etal 1995) 
approves this assumption. Then for the most part of the pulsar orbit the 
geometry of the problem reduces to the interaction of the two spherically 
symmetric winds.

The interaction of the relativistic and non-relativistic winds is still
poor investi\-gated but it seems that it is closely analogous to the case
of the interaction of two non-relativistic winds (e.g. Melatos \etal 1995).
For the non-relativistic case it was
shown  by Lebedev \& Myasnikov (1990) that if the dynamical pressures of the two winds are of the same order then  the form of the shocks converges to a hollow cone at
distances bigger then the distance between the pulsar and a contact
surface.

Equating the dynamical pressures
of the winds we find under the assumption of the winds isotropy the
distance to the contact surface from the pulsar:
  \beq
r_c=a\frac{\sqrt{\a}}{1+\sqrt{\a}}\;\;\;\;\; , \;\;\;\;\;\;
\a=\frac{L_p}{cv_0\dot{M}},
 \label{rc}
 \eeq
 $a$ is a
binary separation, $\dot{M}$ is a Be star mass outflow rate, $v_0$ is a Be
star polar wind termination velocity, $c$ is a light velocity.  In the paper 
of Snow (1982) it was found out that the rate of the Be star mass loss 
is connected with its luminosity. From that dependence it follows that 
the rate of SS2883 mass loss should be about $\dot{M}=$10$^{-8}\dot{M}_{-8}M_{\odot
}$/year. The typical terminal velocity of the Be star polar wind is about 
 $v_0=10^8v_8$cm/c (Snow 1981). Thus  we find that for \b12
$\alpha=0.45/v_8\dot{M}_{-8}$. Hence the winds have comparable values of
the dynamical pressure and the contact surface crosses the binary axis
somewhat closer to the pulsar then to the Be star.

In the beginning both the relativistic and non-relativistic winds are
radial but after passing the shock winds turn and start to flow along
the contact surface.
Both winds are highly supersonic and thus both shocks
arising in the system are strong
and MHD shock conditions states  that the normal component of the particles
velocity after the shock is lowered by a factor 3 and 4 for relativistic
and non-relativistic particles correspondingly (Landau\&Lifshitz 1986).
The big difference in the velocities of the winds at
different sides of the contact surface can lead to the growth of
instability and the two winds will be macroscopically mixed between the
shocks. In this case the region between the shocks will be filled with the
two-phase outflow composed of volumes with relativistic and volumes with
non-relativistic particles.  Then the heavy non-relativistic wind slows
down the volumes filled by the relativistic electrons and positrons and
they acquire essentially non-relativistic hydrodynamic drift velocity $v_d$
along the shock while the energy of electrons and positrons does not
changes significantly. The non-relativistic plasma in its turn accelerates and acquire the velocity of the order of $v_d$. The analysis of the
observational data leads to the conclusion that the similar
situation occurs in systems G70.7+1.2
(Kulkarni \etal 1992) and LSI 61\deg303 (Maraschi\&Treves 1981) and
that the observed emission from these systems is generated
in the regions filled with the two-phase flow.

\section{Generation of the non-pulsed X-ray emission}
The observed power-law spectrum of the non-pulsed X-ray emission coming
from the system \b12 suggests a power-law energy distribution of the
relati\-vistic particles in the outflow after the shock
$\frac{dN_{e^\pm}}{ d\varepsilon }=K_e\varepsilon ^{-s}$,
$\varepsilon>\varepsilon_{min}$
(Chernyakova \& Illarionov 1999).
Such a distribution can be either the result of the intrinsic power law
distribution of the electrons and positrons in the pulsar wind (e.g.
Tademaru 1973; Lominadze \etal 1983), or in the case of the particle
acceleration at the shock (e.g.  Gallant \etal 1994).

The soft photons, emitting by the Be-star would scatter on the relativistic
particles, outflowing with the drift velocity $v_d$.
The scattered photons form a wide spectrum from the X-ray band up to the
gamma-ray band. Applying the results of the work (Chernyakova \& Illarionov
1999)\footnote{In that work there is a misprint - in (27) the value of
 $v_d=10^9$cm/s was taken.} to the system PSR B1259-63  we estimate that for $\varepsilon
_{min}=10mc^2, s=2.4, v_d=10^9$cm/s
 during the periastron passage 
the photon flux density of the unpulsed
X-ray emission from the system at Earth should be about
\begin{equation}
F\sim 5\times 10^{-3}(\frac \varepsilon {100keV})^{-1.7}ph/s/cm^2/MeV,
\end{equation}
while the observations give at 100 keV the value\\ $F_{obs}=(2.8\pm 0.7)\times
10^{-3}$ph/s/cm$^2$/MeV (Grove \etal 1995).

\section{Non-pulsed radio emission as a result of the synchrotron emission
of the electrons of a pulsar wind}
Under the assumption of the relativistic particles power law energy distribution
$dn_{e^\pm}=K_e\varepsilon^{-2.4}d\varepsilon$,
  $\varepsilon>10mc^2$, (the
 parameter  $K_e$ characterize the density of relativistic particles),
 the flux
density detected at the Earth from the synchrotron emission on a frequency
$\nu$ ( throughout the text $\nu$ is measured in GHz) from the small volume
$dV=dldS=dV_{39}*10^{39}$cm$^3$ (with the side $dl$ along the line of
sight) with a magnetic field $B$ is equal to (Corchak, Terleckiy 1952)
\begin{equation}
dI_E=\frac{J_\nu dV}{4\pi D^2}=3.23\times
10^5\frac{K_e}{4\pi}\frac{dV_{39}}{D_2^2}(B\sin{\chi})^{1.7}\nu^{-0.7}mJy,
\label{inu}
\end{equation}
where $\chi$
is an angle between the magnetic field and the direction to
the observer, $D=D_2*2$kpc is a distance to the system and $J_\nu$ characterizes the energy emitted from a unit volume per a unit time.

The coefficient of self-absorption is equal to (Ginzburg \etal 1969)
\begin{equation}
\mu=1.1\times 10^{-8}K_eB^{2.2}\nu^{-3.2}cm^{-1}
\label{pogl}
\end{equation}

In order to calculate the flux density with formulas (\ref{inu}),
(\ref{pogl}) it is necessary to know the density of the relativistic
particles in the outflow $K_e$ and the magnetic field $B$.
To calculate the parameters of the relativistic outflow at an arbitrary
point $I$ located between the two shocks we use a simplifying assumption 
of the parameters constancy across the
shock (along the surface $U$, Fig.1).

\subsection{The relativistic particles density}
 The density of the relativistic particles at the surface
$U$ can be estimated from energy conservation law. Let us consider
contour $ABCDE$ (Fig.1). The energy flow through the closed surface arising
from the rotation of the contour $ABCDE$ about the binary axis is equal to
zero.  Equating the inflowing flux density $L_p\Omega/4\pi$ to
outflowing one $K_eS_Uv_d/0.4\epsilon_{min}^{0.4}$ we find the parameter $K_e$:
  \begin{equation} K_e=\frac{0.4
L_p\Omega\epsilon_{min}^{0.4}}{4\pi S_Uv_d},
 \label{ke}
 \end{equation}
 here $S_U$ is
the surface $U$ area and $\Omega=2\pi(1-\cos\zeta)$, where
$\zeta$ is an angle $\widehat{APC}$ (Fig 1). Far from the pulsar 
$S_U=2\pi R_U^2(\cos\theta_1-\cos\theta_2)$, 
$R_U$- is a distance from the Be star to the surface $U$.

\subsection{Magnetic field}
At distances much bigger then the radius of the light cylinder  the
magnetic field in the pulsar wind may be considered as a toroidal one.
From (\ref{b0}) it follows that at a distance $r$ from the pulsar
it  is equal to $B_w=\frac{1}{r}\sqrt{\frac{\sigma L_p}{c}}$. To calculate
the structure of a magnetic field in the outflow we use the equation of the
frozen magnetic field that in the stationary case have form (Lundquist
1951) \begin{equation} rot[\vec{v} \times \vec{B}]=0. \label{frozen}
 \end{equation}
Let us introduce the cylindrical coordinate system ($\eta,\varphi,z$) with
 $z$ along the binary axis, chosen so that $\varphi_{obs}=0$.
 Assuming that the outflow is
azimuthal symmetrical about the binary axis, that is $v_\varphi=0$,
we find from (\ref{frozen}):
  \begin{equation}
B_{\varphi}=\sqrt{\sigma cL_p}
\frac{\zeta\sqrt{\cos^2\a+\sin^2\a\cos^2(\varphi-\varphi_p)}}
{v_d\int^{\theta_2}_{\theta_1}\frac{\eta_U}{\sin{\theta}}d\theta},
\label{b2} \end{equation} the integration in the denominator is done along
 the surface $U$ at a constant $\varphi$, $\eta_U$ is a distance between
the elements of the surface $U$ and the binary axis,
 vector $\vec{e_p}$ with the coordinates
$(\sin\a,\varphi_p,\cos\a)$ in our cylindrical coordinate system is aligned
with the pulsar axis.

In the degenerate case of the coincidence of the binary axis and the axis
 of the pulsar ($\alpha=0$) the magnetic field has only $\varphi$-component.

In general case the magnetic field in the outflow also have  a component
$B_d$ along the drift velocity and a component $B_U$ along the surface $U$.
The condition of the frozen
 magnetic field suppose that the variation of $B/n_e$ at a volume with a
given particles in a course of motion is proportional to the stretching of a
magnetic field line passing through this volume (Lundquist 1951). With the
 help of (\ref {ke}) we find that $B_\varphi$ and $B_U$ are proportional to
$r^{-1}$ and that $B_d\sim r^{-2}$. However if the mixing of the two winds
takes place then the direction of the component perpendicular to
$\vec{e}_\varphi$ changes all the time in the course of motion, this
component damps faster then the $B_\varphi$ and at a large distance from
the pulsar the $\varphi$-component of the magnetic field dominates.

If the mixing of the wind doesn't occur, then far from the pulsar,
 in the
 region where the form of the shock waves $S_1$, $S_2$ is approximately a
 cone, $B_d$ is small and we find from (\ref{frozen}) the value of $B_U$:
  \beq
 B_U=
 -\frac{\sqrt{\s cL_p}}{\sin{\th_2}-\sin{\th_1}}\frac{\zeta}{R_U v_d}
\frac{\sin^2\a\cos(\varphi-\varphi_p)\sin(\varphi-\varphi_p)}
{\sqrt{\cos^2\a+\sin^2\a\cos^2(\varphi-\varphi_p)}}
\label{bth}
\eeq

\subsubsection{The intensity of the non-pulsed radio emission}
With the help of formulas (\ref{inu}), (\ref{pogl}), (\ref{ke}),
(\ref{b2}) and (\ref{bth}) we can calculate the resulting synchrotron emission
 from the
system. The differential equation describing the change of the flux
density $I_\nu$ along the line of view crossing the emitting region is
\beq
\frac{dI_\nu}{dl}=J_\nu(l)-I_\nu(l)\mu(l),
\label{deq}
\eeq
where $l$ is a coordinate along the line of sight. The total emission that
 would be observed at the Earth  is equal to:
 \beq 
I_E=\frac{1}{4\pi D^2}
\int\int_{l_0}^{l_e}J_{\nu}(l)\exp\left[-\int_{l}^{l_e}\mu(l_1)dl_1\right]
dldS.
\label{ieq}
\eeq
To calculate the observed flux we should choose the limits of integration
$l_0$ and $l_e$ so that this intercept totally covers that part of the line
of sight where the emission is generated  and to integrate over the $l$
and over the area in the plane perpendicular to the line of sight
chosen so that it embraces all the lines of sight crossing the emission
region.

 Figure 2 represents the spectra calculated for different parameters in the case
 of a pulsar with $\alpha=0$ located at a position 
30 days before the periastron passage. In our
calculations we follow the work of Melatos \etal (1995) and approximate the
form of the shock waves $S_1$, $S_2$ with hyperboloids with half-opening angles
$\theta_1, \theta_2$.  It turns out that the bulk of radiation comes from
the regions away from the pulsar, where the drift velocity depends only
slightly on the position along the layer and for the simplicity we treat it
as a constant.  The separation of the region generating the bulk of the
observed emission from the pulsar also leads to the weak dependence of the
result on the form of the shock waves in the vicinity of the pulsar.
For the comparison the observational spectra that were received by Johnston \etal (1999) 3 days before the periastron passage and 16 days after are also shown.

\begin{figure}[t]  
\epsfig{file=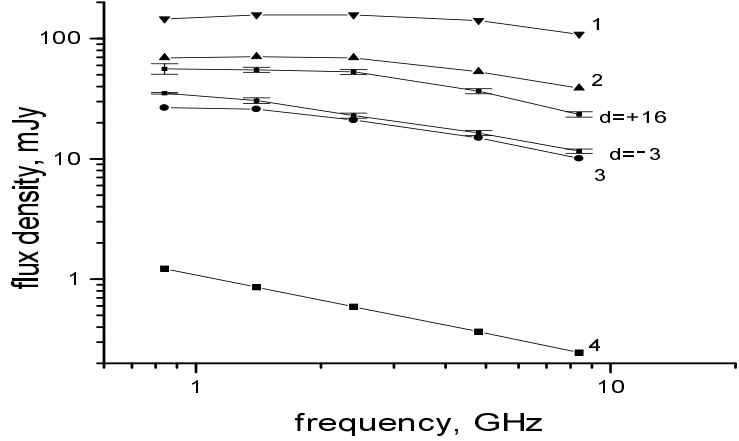,width=12cm, height=12cm}
\vspace{-5cm}
\caption{ The model spectra of the unpulsed radio emission coming from
 the system 30 days from the periastron. In calculations the pulsar axis
 was assumed to be directed along the binary axis.
1.$v_d=c/15$, $\sigma=5\times 10^{-3}$, $\th_1=10\degm$, $\th_2=70\degm$
2.$v_d=c/10$, $\sigma=5\times 10^{-3}$, $\th_1=10\degm$, $\th_2=70\degm$
3.$v_d=c/10$, $\sigma=10^{-3}$, $\th_1=10\degm$, $\th_2=70\degm$
4.$v_d=c/3$, $\sigma= 10^{-3}$, $\th_1=20\degm$, $\th_2=55\degm$.
Curves with error bars are the spectra observed from the system 3 days
 before the periastron and 16 days after (Johnston \etal 1999).}
\end{figure}

Figure 2 shows that the flux density of the emission highly depends on the
 value of the drift velocity $v_d$ and the magnetization parameter
 $\sigma$.The lower the drift velocity is and the higher is the value of
 $\sigma$ the more intensive is the resulted emission.

In the case of the arbitrary position of the pulsar axis the estimation of the
value of the magnetic field along the layer is rather difficult. However as
we saw in the previous section if the mixing occurs then  at large distances
 from the pulsar the $\varphi$-component of the magnetic field dominates and
 its value is close to the one in the degenerate case. The wide range of
the results received in the case of a pulsar with an axis along the binary
axis by variation of parameters shows that observed radio emission can
be explained with a synchrotron mechanism.

\subsubsection{Polarization properties}
The degree of polarization $\Pi$ of the relativistic electrons  with
the power-law energy distribution
$dn_{e^\pm}=K_e\varepsilon^{-2.4}d\varepsilon$
is given by (Corchak, Terleckiy 1952):
 \beq
\Pi=\frac{\sqrt{{Q}_E^2+{U}_E^2}}{{I}_E}
\label{pol}
\eeq
Here $I_E$ is given by (\ref{ieq}) and the formulas for $Q_E$ and $U_E$ can
be received from (\ref{ieq}) by  a substitution of
$0.72J_\nu(l)\cos 2\tilde{\chi}$  and $0.72J_\nu(l)\sin 2\tilde{\chi}$
for $J_\nu(l)$ correspondingly. $\tilde{\chi}$ is an angle between the arbitrary fixed
direction in the plane perpendicular to the line of sight and the direction
along the $\vec{e}_1=\vec{B}\times\vec{e}_o$, where $\vec{e}_o$ is a unit
vector along the line of sight.

On Figure 3 the dependence of the degree of polarization on the angle
between the line of sight and the binary axis is shown. This dependence was
calculated according to (\ref{pol}) for the different positions of the
pulsar axis. It can be seen from the Figure 3 that the observed unpulsed radio emission should be highly polarized. However in formula (\ref{pol})
the effect of the Faraday rotation was not taken into account. The emission
is totally depolarized if the strong Faraday rotation is taken place in the
emission region (Burn, 1966):  \beq
\Delta\varpi=\frac{0.236}{\nu^2}\int nB\cos\chi dl_{13}\gg 1.
\label{depol} \eeq Here $\nu$ is measured in GHz, $B$ in Gs , $n$ in
 cm$^{-3}$ and $l_{13}$ in 10$^{13}$cm.

If beyond the shock wave the winds stirring takes place, then the
electrons density in (\ref{depol}) is the density  of
non-relavistic electrons $n_{nr}$ and the magnetic field is a magnetic
field of the Be-star $B_{Be}$. In the outflow $B_{Be}$ is about 1G and
$n_{nr}$ can be estimated from the particles conservation law
 \beq
n_{nr}=\frac{\dot{M}(1-\cos\th_2)}{4\pi m_p v_{nr} R_U^2 (\cos\theta_1-cos\theta_2)}= 3\times 10^5\frac{\dot{M}_{-8}(1-\cos\th_2)}{v_{nr9}R_{U13}^2(\cos\theta_1-cos\theta_2)}cm^{-3}.
\label{nnr}
\eeq
Here $v_{nr}=v_{nr9}10^{9}$cm/s is a drift velocity of the non-relativistic
plasma beyond the shock wave and $\dot{M}$= $\dot{M}_{-8}M_\odot$/year is the rate of Be star mass
loss. $R_U=R_{U13}10^{13}$cm is a distance from the Be-star to the surface U.
The value of $n_{nr}$ varies along the shock wave but in the region of
 interest it is, as it can be seen from (\ref{nnr}), high and depolarization takes place.  If the mixing
doesn't occur and shock waves have such a form that the line of
sight doesn't pass through the region filled with the non-relativistic
electrons then from (\ref{ke}), (\ref{b2}), (\ref{bth}) and  (\ref{depol})
it follows that $\Delta\varpi\sim1$, the effect of the Faraday rotation is
not strong and the radio emission would be polarized. Unfortunately
the degree of polarization in this case is not measured yet  as the
intensity of the non-pulsed radiation in this case is low (see
Fig.2).

\begin{figure}[t]  
\centerline{\epsfig{file=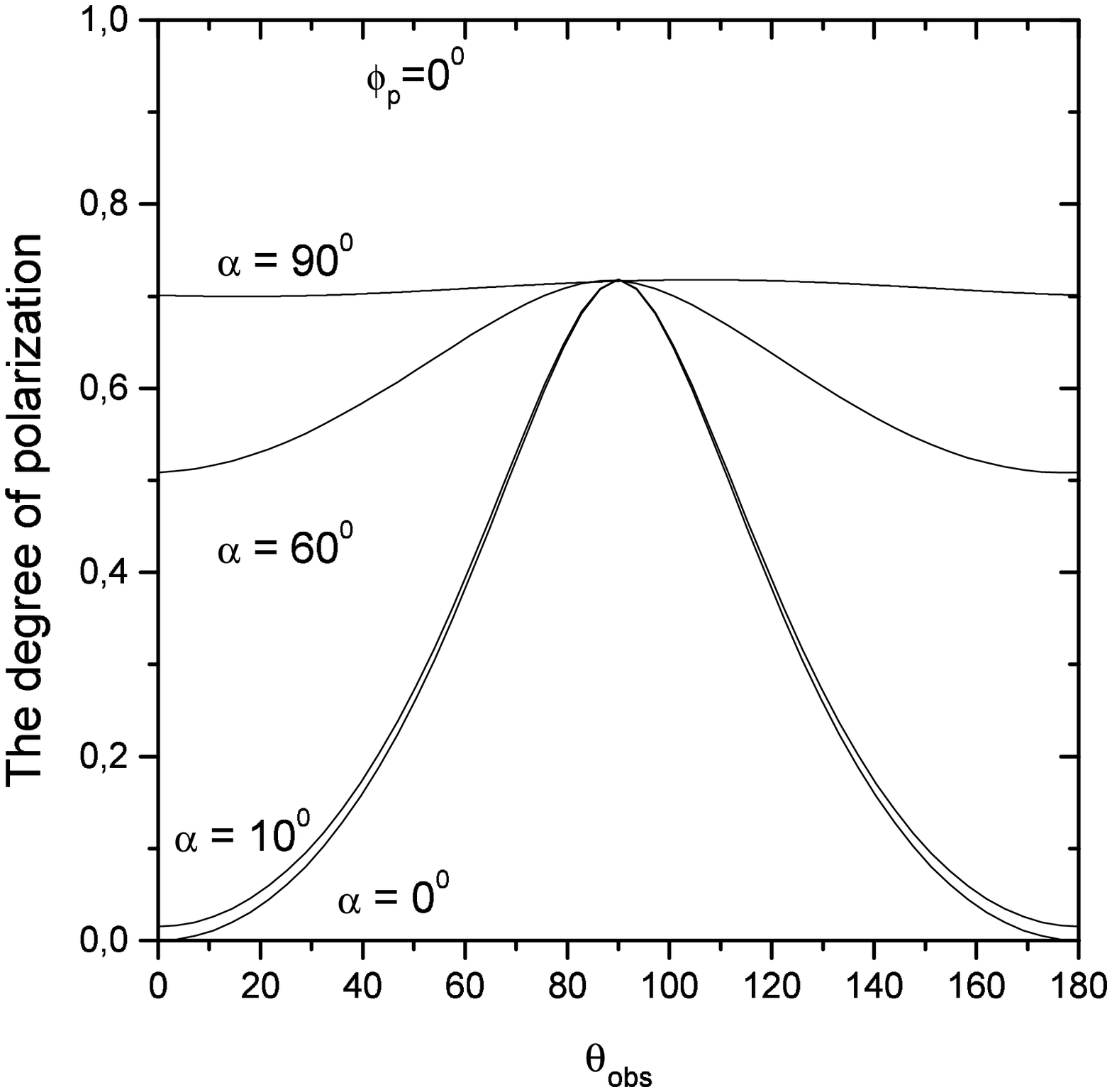,width=6cm, height=9cm}
            \epsfig{file=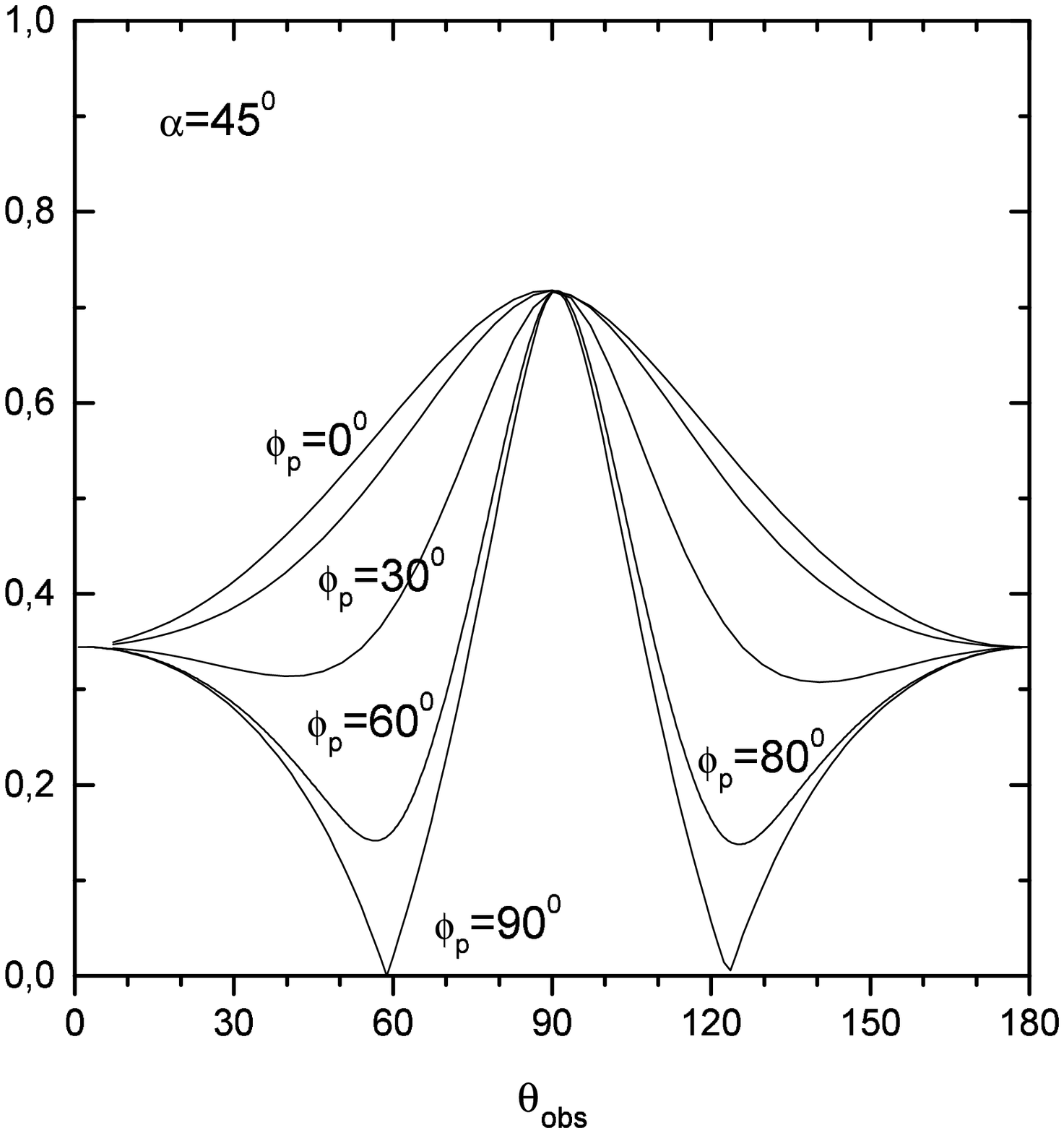,width=6cm, height=9cm}}
\vspace{-2cm}
\caption{ The dependence of the degree of polarization on
the angle $\theta_{obs}$ between the line of sight and the binary axis. Angles
$\alpha$ and $\varphi_p$ define the position of the pulsar axis relative to the binary axis and the observer (see text). It was
calculated
under the assumption that the effect of the Faraday rotation is
negligible.}
\end{figure}

\section{The light curve}
The observations of the \b12 show that far from the periastron there is
no evidence for any
unpulsed radiation in the off-pulse bins down to a limit of several mJy.
The examination of the synchrotron emission of the pulsar wind relativistic
particles in the outflow beyond the shock leads us to the conclusion that
the drift velocity of the relativistic particles is high and there is no
mixing of the relativistic and non-relativistic winds beyond the shock.
According to our analysis the arising unpulsed radio emission is polarized
with the degree of the polarization about 30\%.

The
detectable non-pulsed
radio emission from the system appears only 20 days before the
periastron passage and approximately at the same time the
pulsed emission disappears. It testifies that at this time the pulsar
crosses the disk of Be-star first time and the pulsed emission is absorbed
by the matter of a disk. The non-pulsed radio emission is generated rather
far from the pulsar and crosses the disk at its low-density part. Thus the
free-free absorption of the non-pulsed emission by a disk matter is
negligible.

The interaction with the pulsar  leads to the partial destruction of the
Be star disk (Ivanov \etal 1998). The matter ejecting from the disk
increases the instability in the outflow beyond the shocks leading to the
mixing of the relativistic and non-relativistic winds and to the
deceleration of the relativistic particles drift velocity. As it can be
seen from the Figure 2 the decreasing of the drift velocity leads to the
increase of the intensity of the generated non-pulsed radio emission. The
observed intensity can be explained with the drift velocity about  $v_d\sim
c/10.$ The mixing of the winds leads to the increase of the electron
density on the line of sight and the emission becomes unpolarized.
The free-free optical depth along the line of sight is given by
$\tau_{ff}=5.7\times 10^{-15}\nu^{-2}T_{e5}^{-3/2}\int n_e^2dl_{13}$ ($\nu$
is measured in GHz, $l_{13}$ in 10$^{13}$cm, $n_e$ in cm$^{-3}$ and $T_{e5}$ in
$10^{5}$K). From (\ref{nnr}) we see that $\tau_{ff}\ll 1$

The observed maximum of the unpulsed radio emission intensity
occurs at the time of the second pulsar intersection with the Be
star disk. After that the  ejection
of the matter from the disk continues for some time but the father the
pulsar is from the disk the less is the influence of the ejected matter on
the winds outflow properties and at some moment the winds are not mixing
anymore. The unpulsed radio emission becomes again polarized
but weak.

By this means we show that the observed properties of the non-pulsed
radio\-emission from the system \b12 can be explained by the synchrotron
emission of the pulsar wind relativistic particles in the outflow beyond
the shock wave and that the non-pulsed X-ray emission is due to the
inverse Compton scattering of the Be star photons on this
particles.  }

\acknowledgements 
Authors are grateful to A.V.Myasnikov for the helpful remarks.
We acknowledge the valuable comments of the anonymous referee.
This work was supported by RFBR grant
97-02-16975

 \end{document}